\pdfoutput=1
\documentclass[11pt]{article}
\usepackage{amsmath,amssymb}
\usepackage{newtxtext,newtxmath}
\usepackage{graphicx}
\usepackage[letterpaper,margin=1in]{geometry}
\date{}
\usepackage{cite}
\usepackage{url}
\usepackage[colorlinks=true,linkcolor=blue!60!black,citecolor=blue!60!black,urlcolor=blue!60!black]{hyperref}
\usepackage{multirow,booktabs,longtable,array,float,xcolor}
\usepackage[version=4]{mhchem}
\usepackage{soul,enumitem}
\usepackage[labelfont=bf,font=footnotesize,skip=4pt]{caption}

\usepackage{tabularx}
\AtBeginDocument{}

\title{\bfseries\boldmath When Is Magnetic Information Needed in Machine-Learning Interatomic Potentials?}
\author{\parbox{0.95\textwidth}{\centering
Yanhao Deng$^{1}$, Cheuk Hin Ho$^{2}$, Mengke Li$^{1}$, Lingxiao Li$^{1}$, \\ Bonan Zhu$^{3,4\ast}$, Zeyu Deng$^{1\ast}$\\[8pt]
\small
$^{1}$Department of Materials Science and Engineering, National University of Singapore, 9 Engineering Drive 1, Singapore 117575, Singapore\\
$^{2}$Independent Researcher\\
$^{3}$School of Aerospace Engineering, Beijing Institute of Technology, Beijing, China\\
$^{4}$State Key Laboratory of Environment Characteristics and Effects for Near-space, Beijing, China\\[6pt]
$^{\ast}$Corresponding authors. Email: bzhu@bit.edu.cn; msedz@nus.edu.sg
}}

\begin{document}
\maketitle
\begin{abstract}

Machine-learning interatomic potentials (MLIPs) bring first-principles accuracy to large-scale simulation. Most predict energies and forces from the atomic structure alone. In magnetic materials, nearly identical atomic structures can adopt different magnetic states with different forces, which geometry-only models cannot distinguish. We introduce a descriptor that measures this magnetic ambiguity from the training data. It ranks the gain from magnetic information across five transition-metal datasets. In a general-purpose dataset of 84 elements, the high-ambiguity elements account for more than half of the atoms outside the lanthanides. The ambiguity also makes reference forces sensitive to calculation settings, so lower fitting error does not mean more reliable data. Geometry-only models still reproduce ion migration and vibrations, partly through force-error cancellation within the host structure, but reproduce magnetic-state-dependent structures less well. Magnetic ambiguity is therefore common in current training data, and screening for it before training guides model choice for each target property.\end{abstract}

\section{Introduction}

Machine-learning interatomic potentials (MLIPs) reach near first-principles accuracy at a cost low enough for large-scale and long-timescale molecular dynamics \cite{chen_universal_2022,deng_chgnet_2023,NEURIPS2022_4a36c3c5,musaelian_learning_2023,fu_learning_2025}.
Pre-trained models now extend this capability across much of the periodic table \cite{batatia_foundation_2025,jacobs_practical_2025,yuan_foundation_2026},
and their accuracy is set largely by the density functional theory (DFT) datasets they learn from, such as MPtrj \cite{deng_chgnet_2023}, MatPES \cite{kaplan_foundational_2025} and OMat24 \cite{barroso-luque_open_2026}.
Spin-polarized calculations yield local magnetic moments alongside energies and forces, but not every dataset stores them.
Most MLIPs take only atomic species and positions as input.
The groups building the next datasets and models therefore face a practical decision for magnetic compounds: whether magnetic degrees of freedom need to be stored and learned.

The decision matters because a geometry-only model assigns a single energy and force to each atomic configuration.
In magnetic materials, DFT can converge to several self-consistent solutions at the same geometry, which differ in local moments, spin states, or orbital occupations \cite{ponet_energy_2024}.
Hubbard $U$ corrections reduce self-interaction error for localized $d$ states by favoring near-integer occupations, and in doing so can stabilize additional solutions \cite{ponet_energy_2024,dudarev_electron-energy-loss_1998}.
The reference forces are then not a single-valued function of geometry, and a geometry-only MLIP can fit only one response for states it cannot tell apart.
Benchmarks across metallic elements already indicate that difficulty in learning forces can originate in the electronic structure rather than in model capacity or structural coverage \cite{geng_origin_2026}.

Several approaches have been developed to incorporate electronic-state information into MLIPs.
One strategy distinguishes different oxidation states of the same element as different atomic species \cite{malica_teaching_2025,petersen_importance_2026}.
Models such as CHGNet \cite{deng_chgnet_2023} and charge-aware PaiNN (cPaiNN) \cite{petersen_limitations_2026} instead learn magnetic moments or atomic charges from the embedded atomic representations, while the energy remains a function of atomic species and geometry.
A third class treats local magnetic moments as explicit input degrees of freedom alongside atomic coordinates, using either collinear formulations, such as magnetic moment tensor potentials \cite{novikov_magnetic_2022} and spin-informed universal graph neural networks \cite{xu_spin-informed_2025}, or non-collinear formulations that represent moments as vectors, such as MagNet \cite{yuan_equivariant_2024}, the magnetic atomic cluster expansion \cite{rinaldi_non-collinear_2024}, and magnetic MACE (mMACE) \cite{ho_equivariant_2026}.

Explicit magnetic input adds cost at both stages.
Training requires spin-polarized reference data, and each simulation needs initial moments that are then optimized together with the atomic positions.
Whether a dataset justifies this cost cannot yet be judged before training, and validation errors after training give only partial guidance \cite{fu_forces_2023,jacobs_practical_2025,pota_thermal_2026}.
Geometry-only MLIPs with force errors of approximately 0.1--0.2~eV/\AA\ have reproduced lithium diffusion, ion-migration barriers, energy--strain relations and phase-transition trends \cite{ran_dynamic_2025,zhou_elucidating_2025,zhang_origin_2023}, so a large force error alone does not show that a target property will be wrong.

Here we ask when explicit magnetic information improves force prediction and when it matters for physical properties.
Comparing geometry-only MACE \cite{NEURIPS2022_4a36c3c5} with mMACE on transition-metal compounds computed under different Hubbard $U$ and spin-polarization settings, we find that force errors are large where similar local geometries support distinct magnetic states.
We define a training-free descriptor of this magnetic ambiguity from DFT geometries and moments alone, and its ranking matches the force-accuracy gain from magnetic input across five transition-metal datasets.
Applied to MatPES, it identifies high-ambiguity elements that account for more than half of the non-lanthanide sites.
Geometry-only models still reproduce migration barriers and phonons, partly because force errors on framework atoms cancel within each polyhedron, but they describe the equilibrium structures of different magnetic states less well.

Because the descriptor uses only quantities that spin-polarized DFT calculations already compute, it can be applied to existing and planned datasets independently of the MLIP architecture.
The same analysis could be extended to other electronic quantities absent from the model input, such as charge states, orbital occupations and excited states that differ in local spin or charge.
Foundation models trained on such datasets are now used across battery, catalysis and magnetic-materials research, and the descriptor allows the decision on magnetic degrees of freedom to be made before training and matched to the properties a model is meant to predict.
The ambiguity also makes reference forces sensitive to DFT settings, which shows that fitting error alone is not a reliable measure of reference-data quality when the labels depend on such hidden electronic states.
\section{Results}

\subsection{Magnetic ambiguity limits force accuracy}
\begin{figure}[tbp]
\centering
\includegraphics[width=\textwidth]{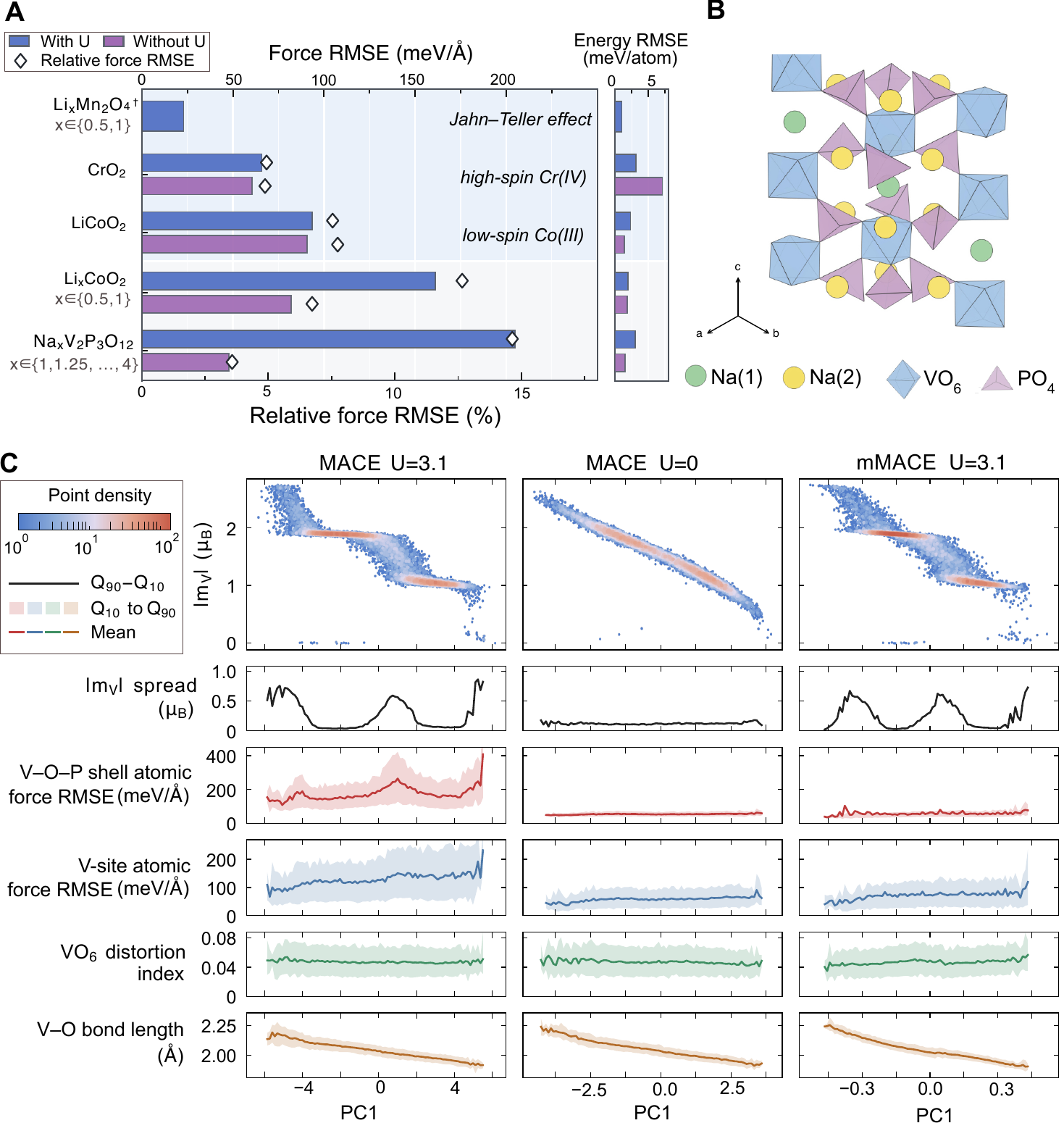}
\caption{\textbf{Material-dependent force errors and their origin in NVP.}
    (\textbf{A}) Energy and force RMSEs of MACE models fine-tuned for transition-metal compounds. Blue and purple bars denote models trained on spin-polarized DFT data with and without a Hubbard $U$ correction, respectively, and diamonds show the force RMSE normalized by the RMS of the DFT force components, expressed as a percentage (bottom axis). $^\dagger$Results from Dillenz et al. \cite{dillenz_interplay_2026}.
    (\textbf{B}) Crystal structure of rhombohedral \ce{Na_xV2(PO4)3}, shown in the crystallographic orientation indicated by the axes at the lower left. The blue \ce{VO6} octahedra and purple \ce{PO4} tetrahedra form a three-dimensional V--O--P polyanionic framework containing interconnected Na-ion channels. Green and yellow spheres denote the crystallographic Na(1) and Na(2) sites, respectively.
    (\textbf{C}) The three columns compare MACE trained on $U=3.1$ eV data, MACE trained on $U=0$ data, and mMACE trained on $U=3.1$ eV data. The top panels show the distribution of V magnetic moment magnitudes, $|m_\mathrm{V}|$, with respect to the PC1 of the corresponding model descriptors, with color indicating the local point density. The lower panels show, along PC1, the spread of $|m_\mathrm{V}|$ defined by the 10\%-90\% range, the force RMSE of V--O--P shell, the V-atom force RMSE, the \ce{VO6} polyhedral distortion index, and the V--O bond length. Shaded regions indicate the 10\%-90\% range of each quantity within local PC1 bins, and solid lines denote the corresponding mean values. }
\label{fig:fig1-summary}
\end{figure}

Force errors of MLIPs fine-tuned with a common workflow differ strongly between transition-metal compounds.
Fig.~\ref{fig:fig1-summary}A compares the energy and force errors of MACE models fine-tuned for several of these compounds, using the dataset construction and fine-tuning workflow of our previous study \cite{deng_accelerating_2025}.
Reference configurations were sampled from finite-temperature trajectories and augmented by strain perturbations, and the number of configurations in each dataset is listed in table~S1 (Methods).
The \ce{Li_xMn2O4} results are taken from Dillenz et al. \cite{dillenz_interplay_2026}, who also fine-tuned MACE-MP on MD-sampled structures, and are included as an external comparison.
Energy errors remain below 5~meV/atom in every system except \ce{CrO2} without $U$, whereas force errors range from less than 50 to approximately 200~meV/\AA.
Because data generation and training are consistent and sufficient for every system, these differences suggest factors beyond the training procedure are at play.

Force errors are small where the relevant magnetic states are either constrained or distinguishable from the local atomic environment.
This is the case for \ce{Li_xMn2O4} (24~meV/\AA), \ce{CrO2} (60.5--65.8~meV/\AA) and \ce{LiCoO2} (90.6--93.5~meV/\AA), with or without a Hubbard $U$ correction.
In \ce{Li_xMn2O4}, Jahn--Teller distortions give the Mn electronic states characteristic local bond-length patterns \cite{dillenz_interplay_2026}.
In \ce{CrO2} and \ce{LiCoO2}, the sampled magnetic states are instead comparatively constrained, with Cr(IV) favoring a high-spin configuration in the ferromagnetic phase and Co(III) predominantly adopting a low-spin configuration \cite{korotin_cro_1998,montoro_electronic_1999}.
The magnetic-moment distributions in our training datasets are consistent with these characteristics (fig.~S1).

The combined \ce{Li_xCoO2} ($x=0.5,1$) dataset and the \ce{Na_xV2(PO4)3} ($x=1,1.25,\dots,4$) composition series (NVP, with \ce{N_xVP} denoting a specific composition) show much larger force errors when the DFT calculations include a Hubbard $U$ correction, with force RMSEs of 161.1 and 204.7~meV/\AA, respectively.
For NVP, this large error persists in models trained separately at individual compositions (table~S1), so it is not simply a consequence of combining different compositions and valence states in one dataset.
Without $U$, the force RMSE falls to 82.0~meV/\AA\ for \ce{Li_xCoO2} and 47.9~meV/\AA\ for NVP.
We also fine-tuned MACE on DFT data generated without spin polarization, which removes spin-dependent variability from the reference calculations.
At $U=0$, this reduces the force RMSE further, to 71.2~meV/\AA\ for \ce{Li_xCoO2} and 33.2~meV/\AA\ for NVP, with the detailed validation performance listed in table~S1.
\ce{CrO2} is excluded from this test because a non-spin-polarized description of a ferromagnetic half-metal is not physically meaningful.
These reductions link the fitting difficulty to spin-dependent variability in the reference data.

To understand the origin of these material-dependent force errors, we take NVP as a representative case and examine the relationship between local geometry, magnetic states, and force prediction under different Hubbard $U$ settings.
As a NaSICON-type cathode, NVP consists of a three-dimensional framework of corner-sharing \ce{VO6} octahedra and \ce{PO4} tetrahedra that forms interconnected Na diffusion channels \cite{deng_fundamental_2022,wang_phase_2022,wang_kinetic_2023}.
In the parent rhombohedral structure ($R\bar{3}c$), Na occupies two crystallographic sites, Na(1) between \ce{VO6} octahedra and Na(2) around the lantern units formed by corner-sharing \ce{VO6} and \ce{PO4} polyhedra (Fig.~\ref{fig:fig1-summary}B) \cite{wang_phase_2022}.
We compare geometry-only MACE with mMACE, which explicitly incorporates local magnetic moments \cite{ho_equivariant_2026}.
The two models share the same message-passing architecture, cutoff and training protocol, and differ mainly in that mMACE additionally takes local magnetic moments as input, so differences in force accuracy reflect the effect of the magnetic input.
Fine-tuned on the same $U=3.1$~eV data with the same hyperparameters, mMACE reduces the validation force RMSE from 204.7 to 85.8~meV/\AA, with an energy RMSE of 1.2~meV/atom. This improvement indicates that missing magnetic information contributes to the force errors of the geometry-only model. The remaining mMACE errors are concentrated in a small population. Excluding the 5\% of atoms with the largest force errors reduces its force RMSE over the complete dataset (including training and validation structures) from 97 to 36~meV/\AA\ (fig.~S2).

If similar local geometries in NVP correspond to distinct magnetic states with different reference forces, a geometry-only model cannot separate them.
To test this, we compare how the learned representations of MACE at $U=0$ and $U=3.1$~eV and of mMACE arrange the V magnetic moments.
For each V atom, we extract the 256-dimensional atom-centered descriptor produced by each model after message passing, perform principal component analysis (PCA) on these descriptors, and plot several properties along the first principal component (PC1) in Fig.~\ref{fig:fig1-summary}C.
PC1 accounts for 90.4\%, 94.2\% and 76.7\% of the descriptor variance for MACE at $U=3.1$~eV, MACE at $U=0$ and mMACE, respectively, and is therefore the dominant direction in each descriptor space.
In all three models, the V--O bond length decreases systematically with increasing PC1, whereas the \ce{VO6} polyhedral distortion index shows no clear dependence on PC1.
PC1 therefore captures a similar dominant variation in the local V environment in all three descriptor spaces, while the second principal component (PC2) is less strongly related to these structural quantities (fig.~S3).
The distribution of the V moment magnitude $|m_\mathrm{V}|$ along PC1, however, differs between the models.
For MACE trained on $U=0$ data, the learned structural representation remains strongly and uniformly correlated with $|m_\mathrm{V}|$, so the moment is largely determined by the local geometry.
For $U=3.1$~eV data, $|m_\mathrm{V}|$ separates into two sharp distributions, consistent with the tendency of the Hubbard correction to favor more localized, near-integer orbital occupations \cite{dudarev_electron-energy-loss_1998}.
Different geometries within each branch map to a nearly fixed moment, but similar local environments in the transition region between the two branches can support distinct moments, giving an effectively non-unique mapping from geometry to magnetic state.

The effect of this ambiguity on the forces extends beyond the V atoms.
Force errors on V atoms alone show only a weak correlation with the moment distribution (Fig.~\ref{fig:fig1-summary}C, lower panel), whereas the force error is more strongly correlated with the spread in $|m_\mathrm{V}|$ when the analysis is extended to the surrounding V--O--P shell.
For each V atom, we define a local shell \(s\) comprising the central V atom, its six nearest O atoms and the P atom nearest to each O atom, giving 13 atoms in total. The force-component RMSE of this shell is
$
\mathrm{RMSE}_s=\sqrt{\frac{1}{3\times13}\sum_{i\in s}\left|\mathbf{F}_i^{M}-\mathbf{F}_i^{\mathrm{DFT}}\right|^2},
$
where $\mathbf{F}_i^{M}$ and $\mathbf{F}_i^{\mathrm{DFT}}$ are the model-predicted and DFT reference forces on atom $i$, respectively.
The ambiguity of the V moments therefore propagates through the connected V--O--P framework and contributes to the force errors on neighboring O and P atoms.
In mMACE, $|m_\mathrm{V}|$ is distributed along PC1 much as in MACE at $U=3.1$~eV, but because the local moments are encoded explicitly, the model does not need to infer a unique magnetic state from geometry alone and can distinguish force labels associated with otherwise similar local environments.
CHGNet performs similarly to geometry-only MACE on the same NVP training set at $U=3.1$~eV. After excluding the 5\% of atoms with the largest force errors for each model, the force RMSEs are 102 and 121~meV/\AA\ for CHGNet and MACE, respectively, compared with 36~meV/\AA\ for mMACE (fig.~S2).
Although CHGNet uses magnetic-moment supervision to regularize its atomic representations, both moments and energies are inferred from atomic species and geometry \cite{deng_chgnet_2023}. It therefore lacks an independent magnetic-state input to resolve the non-unique mapping identified here, where similar local geometries support distinct magnetic states with different forces.

Fixed-geometry DFT calculations confirm that this ambiguity is present in the reference data themselves.
In the ground-state \ce{N3VP} structure with one V atom displaced by 0.25~\AA, calculations initialized with different ferromagnetic and antiferromagnetic moments converge to two solutions that differ by only 0.3~meV/atom in energy but by 29.5~meV/\AA\ in forces (fig.~S4).
Force prediction is therefore accurate when the sampled magnetic states can be distinguished from local geometry and degrades when they cannot, even within the same dataset.
These observations motivate a dataset-level measure of magnetic ambiguity to assess the potential benefit of explicit magnetic information before MLIP training.

\subsection{A training-free measure of magnetic ambiguity}
\begin{figure}[tbp]
\centering
\includegraphics[width=0.8\textwidth]{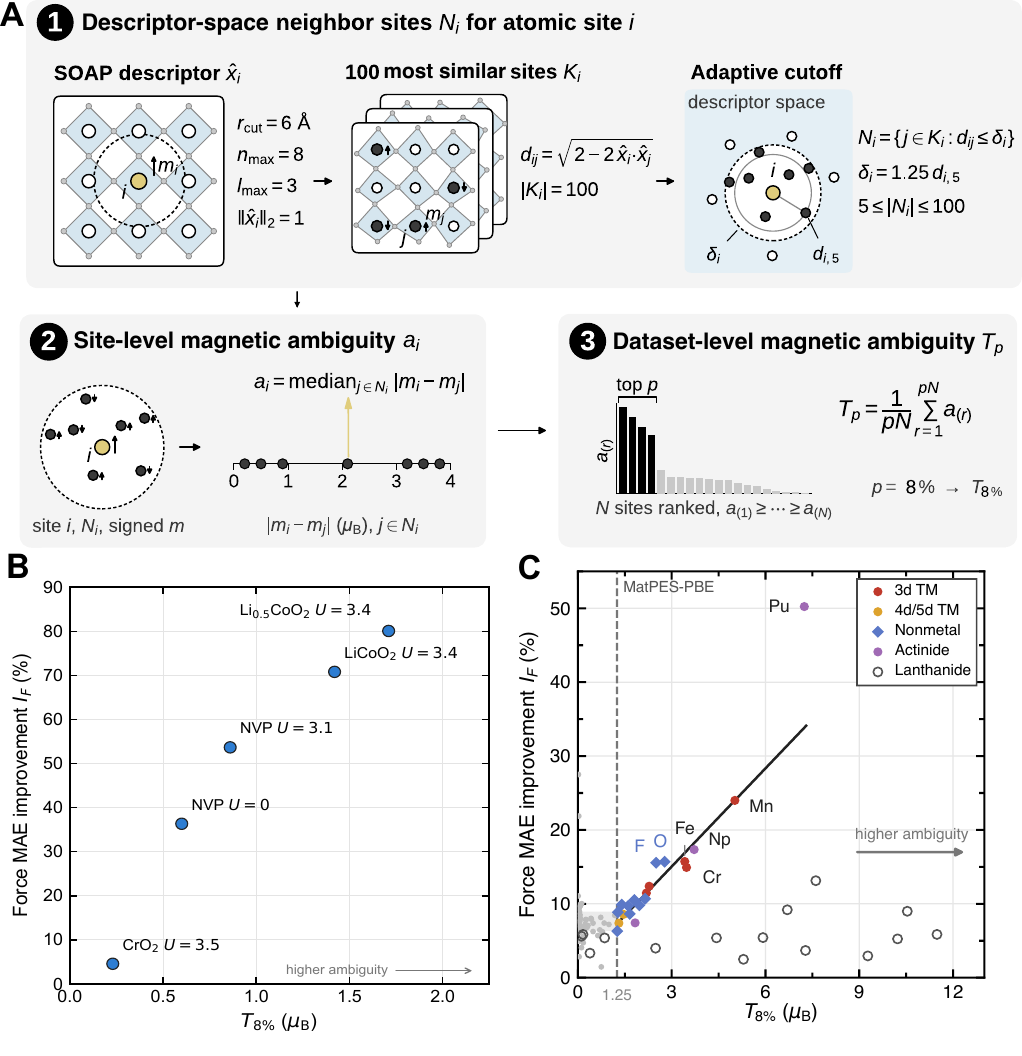}
\caption{\textbf{Dataset magnetic ambiguity $T_{8\%}$ and its relation to the improvement in force accuracy achieved by magnetic models.}
    (\textbf{A}) Workflow to calculate the dataset magnetic ambiguity descriptor $T_p$. 
    (\textbf{B}) Relative improvement in force accuracy from MACE to mMACE, $I_F$, calculated from the force MAEs as
$I_F=\frac{\left(\mathrm{MAE}_{\mathrm{MACE}}-\mathrm{MAE}_{\mathrm{mMACE}}\right)}{\mathrm{MAE}_{\mathrm{MACE}}}\times100\%$,
versus $T_{8\%}$ for the five material-specific datasets. Each MACE--mMACE model pair is fine-tuned from the corresponding pre-trained model.
    (\textbf{C}) $I_F$ versus $T_{8\%}$ for all element populations with at least 1,000 sites in MatPES-PBE. For nonmetals (N, C, P, O, S, Se, F, Cl, Br, I), $T_{8\%}$ is evaluated from the site ambiguities of their directly bonded $d$- and $f$-block atoms. Elements below the threshold $T^{*}=1.25~\mu_\mathrm{B}$ (dashed line) are shown in gray. Lanthanides (open circles) are excluded from the fit. 
}
\label{fig:dataset-ambiguity}
\end{figure}

Because large force errors arise where similar geometries support distinct magnetic states, the benefit of explicit magnetic information can be assessed directly from a DFT dataset, before any MLIP is trained, by quantifying this magnetic ambiguity (Fig.~\ref{fig:dataset-ambiguity}A).

We represent local atomic environments with the smooth overlap of atomic positions (SOAP) descriptor implemented in DScribe \cite{bartok_representing_2013,himanen_dscribe_2020}.
SOAP is a well-established descriptor whose radial and angular expansions share basic ingredients with MACE.
Though the features are not equivalent to the representations learned by MACE, SOAP provides a model-independent representation of local geometry for assessing magnetic ambiguity, with its effectiveness supported by the strong correlations observed across the datasets examined here.
For each atomic site $i$, we take the 100 nearest sites of the same element in SOAP space as candidates and select its neighbors $N_i$ from them with an adaptive cut-off radius.
The site ambiguity $a_i$ is the median absolute difference between the signed reference moment of site $i$ and those of its neighbors, after the signs are aligned within each configuration.
The dataset magnetic ambiguity $T_p$ is the mean site ambiguity over the most ambiguous fraction $p$ of sites ($0<p<100\%$).
This upper-tail average emphasizes environments with poorly resolved magnetic states, which a population-wide mean would dilute.
To measure the benefit of magnetic input, we define the relative improvement in force accuracy, $I_F$, as the reduction in force MAE from MACE to mMACE, since the two models differ mainly in the magnetic input.
The definitions and implementation are detailed in Methods.
The ranking of datasets by $T_p$ is stable over a broad range of $p$, from roughly 5\% to 40\% for the material-specific datasets and to above 80\% for the MatPES transition metals (Ti, V, Cr, Mn, Fe, Co, Ni, Cu, Nb, Mo, Tc, Ru, Rh and Pd), and is similarly insensitive to the neighborhood parameters (fig.~S5).
We use $p=8\%$ for a consistent comparison across these datasets, because it lies inside this stable plateau while avoiding the noisy regime below 5\%. Population-size convergence tests support using at least 1,000 sites (fig.~S6).

Across the five material-specific datasets, $T_{8\%}$ and $I_F$ follow the same ranking (Spearman $\rho=1$, Fig.~\ref{fig:dataset-ambiguity}B), with detailed force MAEs listed in table~S2. The five datasets are \ce{CrO2} ($U=3.5$~eV), \ce{LiCoO2} and \ce{Li_{0.5}CoO2} ($U=3.4$~eV), and NVP at $U=0$ and $U=3.1$~eV.
\ce{CrO2} shows the lowest ambiguity and only a small benefit from magnetic input ($I_F=4.6\%$), consistent with the preferred high-spin Cr(IV) state identified in the preceding analysis.
In \ce{LiCoO2}, mMACE reduces the force MAE by 70.8\%, despite the small overall MACE force error.
Its second-highest $T_{8\%}$ among the five datasets reflects a small population of highly ambiguous Co sites, not widespread ambiguity (fig.~S7).
Most Co sites have moments close to zero, whereas the highly ambiguous population is concentrated around $2~\mu_\mathrm{B}$ and its ambiguity is predominantly related to moment magnitude.
In NVP, the highly ambiguous sites are enriched in the negative-moment region, and the increase in $T_{8\%}$ from $U=0$ to $U=3.1$~eV is predominantly related to moment magnitude (fig.~S7).
The additional amplitude variation between the populations near 1 and $2~\mu_\mathrm{B}$ is consistent with the transition region in Fig.~\ref{fig:fig1-summary}C, where the two moment-amplitude branches overlap in structural space.

Magnetic ambiguity also varies within a single material as the structural sampling changes.
Supplementary fig.~S8 compares \ce{N2VP} and \ce{Co3O4} datasets generated at MD temperatures from 50 to 900~K.
$T_{8\%}$ increases rapidly over 100--200~K in \ce{N2VP}, whereas \ce{Co3O4} retains low ambiguity over a broader temperature range before increasing markedly above approximately 500~K.

\subsection{Magnetic ambiguity across the periodic table}

\begin{figure}[tbp]
\centering
\includegraphics[width=\textwidth]{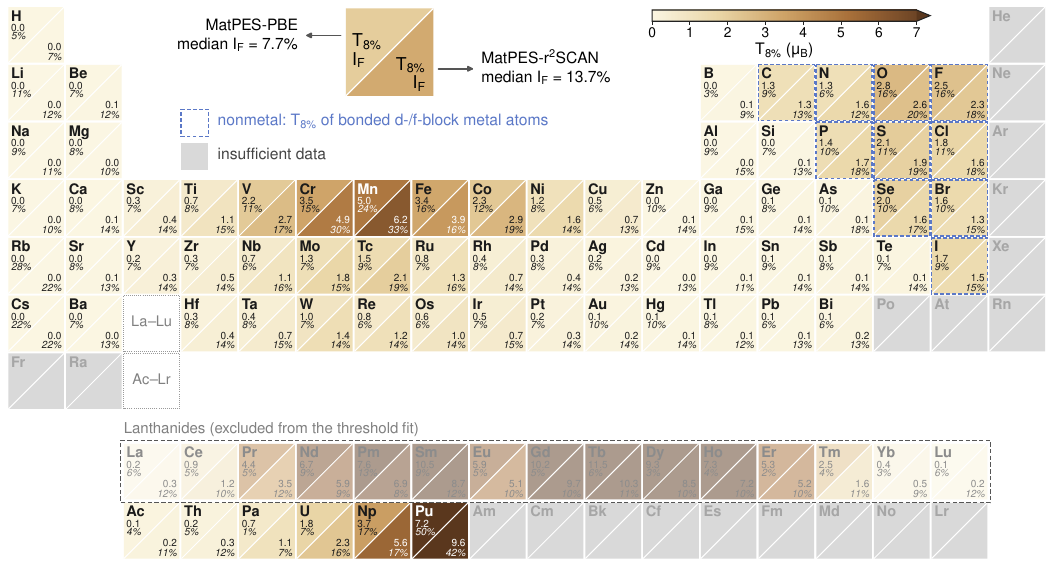}
\caption{\textbf{Element-resolved magnetic ambiguity and force-accuracy improvement in MatPES.}
Each cell is divided along its diagonal, with MatPES-PBE in the upper-left half and MatPES-r$^2$SCAN in the lower-right half.
Colors indicate the magnetic ambiguity $T_{8\%}$ on a common scale, and the numbers in each half give $T_{8\%}$ (in $\mu_\mathrm{B}$) and the force MAE improvement $I_F$ (in \%) from MACE to mMACE.
For nonmetals (dashed blue frames), $T_{8\%}$ is evaluated from the site ambiguities of their directly bonded $d$- and $f$-block atoms, while $I_F$ is evaluated on the nonmetal atoms themselves.
Lanthanides (faded cells) are excluded from the threshold fit (Methods).
Gray cells denote elements with fewer than 1,000 sites or absent from MatPES.
Configurations with reference forces above 20~eV/\AA\ are excluded.
The median $I_F$ in the legend is taken over all 84 evaluated elements for each functional.
}
\label{fig:periodic-table}
\end{figure}

We apply the same analysis to the MatPES-PBE and MatPES-r$^2$SCAN datasets \cite{kaplan_foundational_2025}, which span diverse compounds and local environments under two exchange--correlation functionals.
For each functional, MACE and mMACE are trained from scratch on the full MatPES dataset, not only on compounds containing the selected elements.
All atomic sites of each element are treated as one population, and its magnetic ambiguity and force MAEs are evaluated with the same model pair.
Because $T_p$ is reliable only for populations of at least 1,000 sites (fig.~S6), we include the 84 elements that meet this criterion.
The nonmetals N, C, P, O, S, Se, F, Cl, Br and I carry almost no magnetic moment of their own, so their $T_{8\%}$ is evaluated from the site ambiguities of the directly bonded $d$- and $f$-block atoms, while their $I_F$ is evaluated on the nonmetal atoms themselves. These elements are outlined in blue in Fig.~\ref{fig:periodic-table}.

Elements above an ambiguity threshold account for more than half of the non-lanthanide sites in MatPES.
For most non-lanthanide elements with low magnetic ambiguity, $I_F$ clusters around a dataset-specific baseline. A clearer positive association between $I_F$ and $T_{8\%}$ emerges at higher ambiguity (Fig.~\ref{fig:dataset-ambiguity}C for MatPES-PBE and fig.~S9 for MatPES-r$^2$SCAN).
We describe this pattern with a piecewise fit of a constant baseline and a linear segment, excluding the lanthanides (Methods).
We retain $T^{*}=1.25~\mu_\mathrm{B}$ for MatPES-PBE and use the fitted threshold of $0.93~\mu_\mathrm{B}$ for MatPES-r$^2$SCAN, with median $I_F$ below $T^{*}$ of 7.6\% and 13.5\%, respectively.
Above $T^{*}$, $I_F$ increases with $T_{8\%}$ in both datasets, with Spearman $\rho=0.92$ for PBE and 0.73 for r$^2$SCAN.
Of the 69 non-lanthanide elements, 20 in MatPES-PBE and 28 in MatPES-r$^2$SCAN lie above their thresholds, including several transition metals, actinides, and all ten selected nonmetals.
Although these are a minority of the elements, they account for 56\% of the non-lanthanide sites in MatPES-PBE and 61\% in MatPES-r$^2$SCAN.
Magnetic ambiguity therefore identifies where explicit magnetic input could improve force fitting across a substantial fraction of a large, chemically diverse dataset.
Fig.~\ref{fig:periodic-table} maps $T_{8\%}$ and $I_F$ across the periodic table for both functionals, with element-resolved force MAEs in tables~S3 and S4 and magnetic-moment distributions in fig.~S10.

Among the transition metals, magnetic ambiguity is generally higher for the $3d$ series than for the $4d$ series and decreases down each group.
At group 7, $T_{8\%}$ falls from 5.02~$\mu_\mathrm{B}$ for Mn to 1.46 and 0.78~$\mu_\mathrm{B}$ for Tc and Re in MatPES-PBE.
A likely reason is that the more extended $4d$ and $5d$ orbitals overlap more strongly with their neighbors, which broadens the $d$ bands and increases the ligand-field splitting.
Low-spin or non-magnetic states are then favored, and the local moments are small, with typical mean magnitudes of about 0.1~$\mu_\mathrm{B}$ for the $4d$ and $5d$ metals compared with 0.9~$\mu_\mathrm{B}$ for the $3d$ metals.
Within each series, the ambiguity peaks at group 7 and is generally lower toward either end.
This trend is consistent with the large local moments permitted by nominally near-half-filled $d$ shells.
In an idealized high-spin picture, a $d^n$ configuration permits at most $N_{\mathrm{unpaired}}^{\max}=\min(n,10-n)$ unpaired electrons, reaching five at half filling.
Near-half-filled $d$ shells can therefore support large local moments, which increase the absolute moment differences that arise when structurally similar sites have different moment amplitudes or opposite orientations.
The actual oxidation states, spin states, and bonding environments vary across MatPES.

The ambiguity also depends on the functional, being higher in MatPES-r$^2$SCAN for 66 of the 84 evaluated elements.
One contributing factor may be the larger local moments predicted by r$^2$SCAN \cite{fu_applicability_2018,ekholm_assessing_2018,kothakonda_testing_2023}.
In MatPES, the mean moment magnitude of the $3d$ metals is about 1.5 times as large with r$^2$SCAN as with PBE, and these larger moments amplify the contrast between sites whose moments are aligned with their element's net moment in the configuration and those whose moments are opposed to it (fig.~S11).
For Fe and Np, $T_{8\%}$ increases from PBE to r$^2$SCAN, whereas $I_F$ remains approximately unchanged, illustrating that greater magnetic ambiguity does not necessarily yield a larger force-fitting benefit.
A possible reason is that r$^2$SCAN yields larger and more variable moments for these elements, while the forces respond only weakly to this additional variation.

The lanthanides and the nonmetals O and F show that ambiguity and benefit can decouple in both directions.
The lanthanides show no systematic increase in $I_F$ with their large magnetic ambiguity. In MatPES-PBE, for example, $T_{8\%}$ spans 2.5--11.5~$\mu_\mathrm{B}$ across Pr--Tm, while $I_F$ ranges from 2\% to 13\%, around the low-ambiguity baseline of 7.6\%.
One possible explanation is that the filled $5s$ and $5p$ shells shield the $4f$ electrons and limit their direct participation in bonding \cite{dolg_energy-adjusted_1989,cotton_systematics_2017}, so differences in $4f$ moments may have a limited effect on interatomic forces. 
Pu shows the opposite behavior, combining large ambiguity with the largest gain of any element ($T_{8\%}=7.2~\mu_\mathrm{B}$ and $I_F=50\%$ in MatPES-PBE), as does Np to a lesser extent.
This contrast is consistent with the greater participation of actinide $5f$ electrons in bonding compared with the shielded lanthanide $4f$ electrons \cite{moore_nature_2009}.
O and F, by contrast, carry little local moment but show force MAE reductions above the baseline in both datasets.
This is consistent with the forces on O and F depending on the magnetic states of their bonded metal neighbors through changes in the bonding electronic structure.

Compared with the material-specific datasets in Fig.~\ref{fig:dataset-ambiguity}B, MatPES generally shows larger $T_{8\%}$ but smaller $I_F$.
The scale of $T_{8\%}$ depends on the structural distances between neighbors, which can be larger in a chemically diverse dataset, and $I_F$, as a relative reduction in force MAE, depends on the baseline MACE error.
Comparisons are therefore most informative within a common dataset and training strategy.
Across material-specific and chemically diverse datasets, $T_{8\%}$ connects magnetic-state disagreement between similar local environments to the force-fitting benefit of explicit magnetic information, and it can be evaluated before MLIP training.
The descriptor measures the learnability of the supplied reference data.
Under consistent descriptor and neighborhood settings, lower ambiguity indicates smaller magnetic-moment differences unresolved by the structural representation, not greater physical accuracy of the reference calculations.
Because $T_{8\%}$ captures the moment disagreement of the central sites, the benefit of explicit magnetic input also depends on how strongly the forces respond to magnetic-state differences and on which atoms carry the relevant magnetic information.
The lanthanides and the cross-functional comparisons for Fe and Np show that large or increasing magnetic ambiguity need not translate into commensurate gains in force accuracy, and O and F show how magnetic information from bonded metal neighbors can help predict forces on atoms with small local moments.
For other elements, neighboring moments may also contribute to the forces, although their influence is less apparent alongside that of the central atom's own moment.

\subsection{Effect on physical properties}

\begin{figure}[tbp]
\centering
\includegraphics[width=\textwidth]{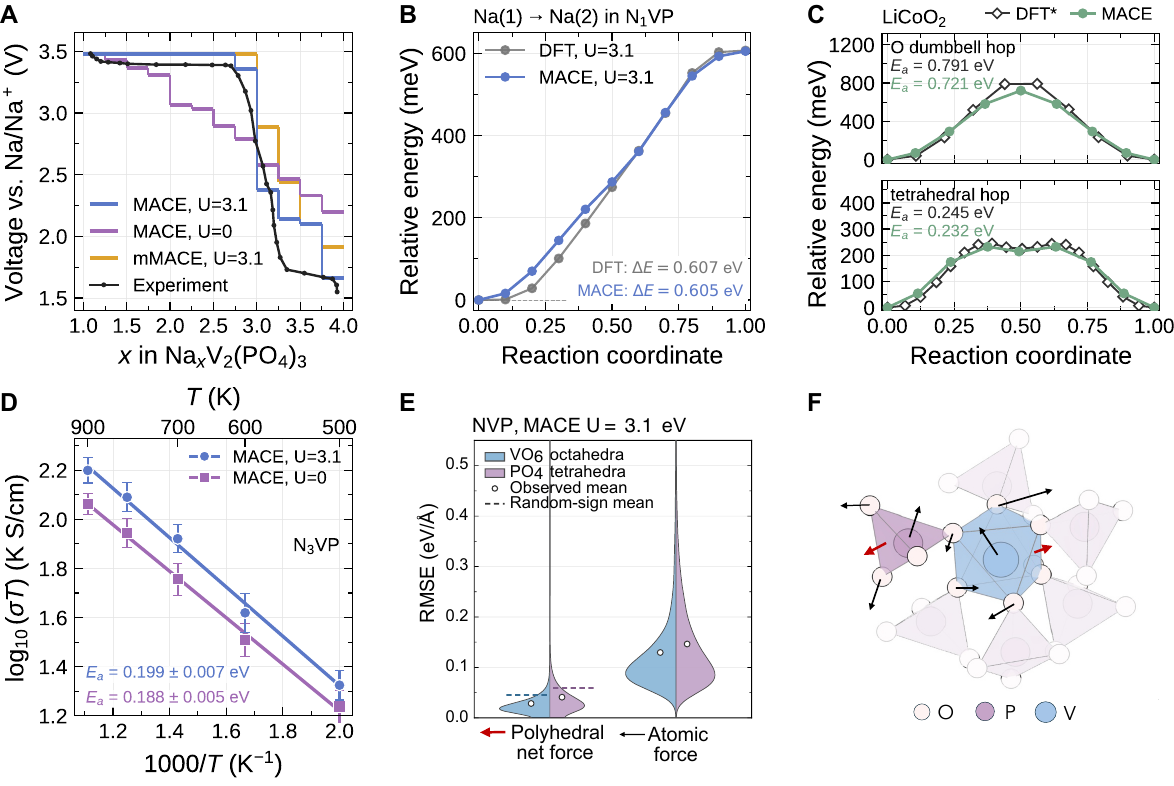}
\caption{\textbf{Physical properties and force-error cancellation in NVP and \ce{Li_xCoO2}.}
(\textbf{A}) Voltage profiles of NVP predicted by MACE at $U=0$ and $U=3.1$~eV and by mMACE at $U=3.1$~eV, compared with experiment \cite{wang_phase_2022}. Each calculated profile is shifted by a constant to match the experimental voltage at the lowest measured Na content.
(\textbf{B}) Relative-energy profiles along the Na(1)$\rightarrow$Na(2) path in \ce{N1VP} from DFT and MACE at $U=3.1$~eV. The plotted profiles rise monotonically. $\Delta E$ denotes the energy difference between their endpoints.
(\textbf{C}) Relative-energy profiles for Li migration in \ce{LiCoO2} along the oxygen-dumbbell pathway, with one Li vacancy, and the tetrahedral-site pathway, with two Li vacancies, from MACE at $U=$~3.4~eV and literature DFT (DFT*) \cite{van_der_ven_first-principles_2001}, with the migration barriers $E_a$ indicated.
(\textbf{D}) Arrhenius plots of the Na-ion conductivity of \ce{N3VP} predicted by MACE at $U=0$ and $U=3.1$~eV. Error bars show the standard deviation for each conductivity point calculated with \texttt{kinisi} \cite{mccluskey_accurate_2025}.
(\textbf{E}) Distributions of polyhedral net force RMSEs and atomic force RMSEs for \ce{VO6} octahedra and \ce{PO4} tetrahedra across the evaluated NVP dataset, from MACE at $U=3.1$~eV. For a polyhedron containing $N$ atoms, these quantities are $|\sum_i\Delta\mathbf{F}_i|/(\sqrt{3}N)$ and $\sqrt{\sum_i|\Delta\mathbf{F}_i|^2/(3N)}$, respectively, where $\Delta\mathbf{F}_i=\mathbf{F}_i^{\mathrm{MACE}}-\mathbf{F}_i^{\mathrm{DFT}}$. White circles indicate means, and dashed lines indicate the means of the random-sign baseline.
(\textbf{F}) Schematic of force-error cancellation within the NVP framework. Black arrows denote atomic force-error vectors, and red arrows show their net vector sums within each polyhedral unit.
}
\label{fig:cathode-property-comparison}
\end{figure}

\begin{figure}[tbp]
\centering
\includegraphics[width=\textwidth]{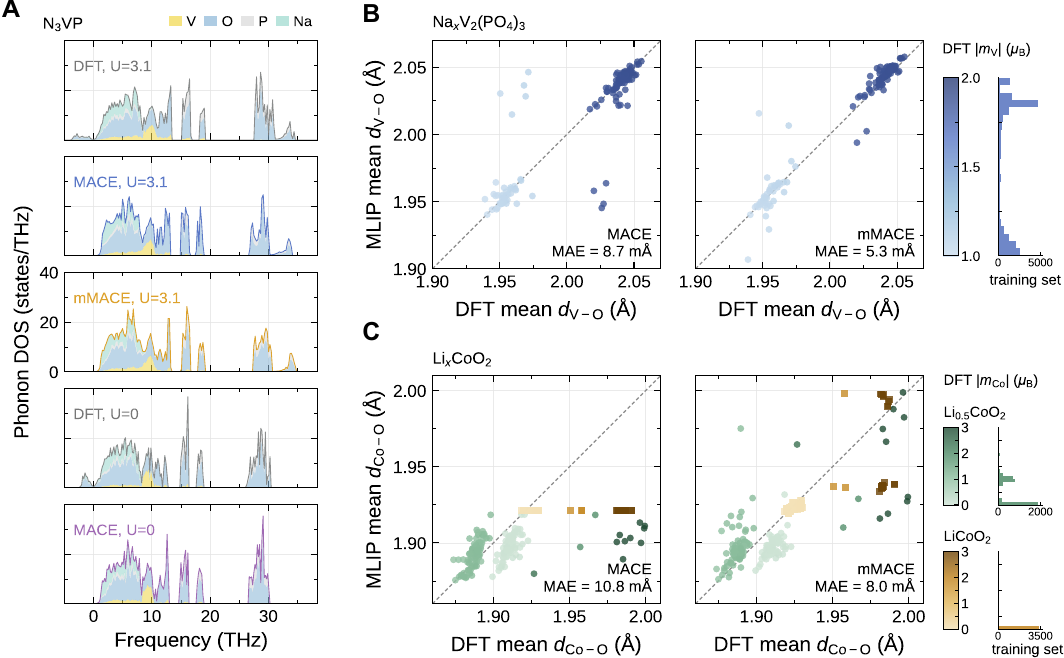}
\caption{\textbf{Phonon densities of states and magnetic-state-dependent local structures after relaxation.}
(\textbf{A}) Element-projected phonon densities of states of \ce{N3VP} from DFT and MACE at $U=3.1$~eV and $U=0$, and from mMACE at $U=3.1$~eV, calculated from finite-displacement force constants using $2\times2\times1$ supercells and an $8\times8\times8$ q-point mesh with the tetrahedron method.
(\textbf{B}) Mean V--O bond lengths in \ce{N_xVP} after MACE (left) and mMACE (right) relaxation against the corresponding DFT-relaxed values at $U=3.1$~eV. Each point is the mean of the six V--O bonds around a V site, colored by its DFT-relaxed magnetic moment magnitude.
(\textbf{C}) Mean Co--O bond lengths in \ce{LiCoO2} (squares) and \ce{Li_{0.5}CoO2} (circles) after MACE (left) and mMACE (right) relaxation against the DFT-relaxed values. In (\textbf{B}) and (\textbf{C}), the adjacent histograms show the magnetic moment distributions in the training data.
}
\label{fig:relaxed-structures}
\end{figure}

The substantial improvement in force accuracy obtained with mMACE raises the question of how strongly the missing magnetic information affects physical properties.
We examine this question using two systems: the NVP models trained jointly across $1\leq x\leq4$ with $\Delta x=0.25$, and the \ce{Li_xCoO2} models trained at $x=0.5$ and 1, with the results shown in Fig. \ref{fig:cathode-property-comparison} and \ref{fig:relaxed-structures}.

Both models trained with $U=3.1$~eV reproduce the main features of the NVP voltage profile, whereas the model trained without $U$ does not, despite its smaller training error (Fig.~\ref{fig:cathode-property-comparison}A).
The profiles are derived from MLIP-calculated formation energies (fig.~S12), and each calculated profile is shifted by a constant voltage offset to match the experimental voltage at the lowest measured Na content.
With $U=3.1$~eV, MACE and mMACE reproduce the broad high-voltage plateau and the subsequent voltage drop at $x=2.75$ and $x=3.0$, respectively.
For $x>3$, both models predict stable structures at each sampled composition on the lower convex hull, producing successive short voltage plateaus instead of the smooth experimental decline.
MACE also reproduces the final low-voltage plateau at approximately 1.7~V versus Na/Na$^+$ near $x=3.75$.
MACE trained on data with $U=0$ instead predicts every sampled composition over $1<x<4$ on the lower convex hull, and the slopes between successive hull vertices yield a descending voltage staircase that is inconsistent with the experimental voltage.
The lower force error at $U=0$ therefore indicates that these reference data are easier for the model to fit, whereas the better agreement with experimental voltages at $U=3.1$~eV shows the importance of the Hubbard $U$ correction for describing the energetics of Na insertion and extraction.

The mobile-ion energy landscapes are well reproduced by geometry-only MACE.
For Na(1)$\rightarrow$Na(2) migration in \ce{N1VP} (Fig.~\ref{fig:cathode-property-comparison}B), MACE at $U=3.1$~eV closely follows the DFT relative-energy profile, with endpoint energy differences of 0.605 and 0.607~eV, respectively.
In \ce{LiCoO2}, MACE retains the main features of the oxygen-dumbbell pathway containing one Li vacancy and the tetrahedral-site pathway containing two Li vacancies, compared with literature DFT values (Fig.~\ref{fig:cathode-property-comparison}C) \cite{van_der_ven_first-principles_2001}.

Finite-temperature Na transport provides a further comparison of the learned energy landscapes.
In \ce{N3VP}, the MACE conductivities at $U=3.1$~eV and $U=0$ remain of the same order of magnitude over 500--900~K (Fig.~\ref{fig:cathode-property-comparison}D), with activation energies of $0.199\pm0.007$ and $0.188\pm0.005$~eV, respectively.
Although the two models describe reference potential-energy surfaces generated with different Hubbard $U$ settings, their substantially different force-fitting errors are not accompanied by comparably large differences in the predicted Na transport.

The phonon density of states (DOS), which probes the local curvature of the potential-energy surface, is also reproduced for its main features (Fig.~\ref{fig:relaxed-structures}A). 
For \ce{N3VP}, the MLIPs reproduce most of the DFT peak frequencies and their dominant elemental contributions over 0--30~THz, and at $U=3.1$~eV both MACE and mMACE also reproduce the high-frequency peaks above 30~THz.
All of the MLIPs remove the shallow imaginary modes present in the corresponding DFT calculations (not shown in Fig.~\ref{fig:relaxed-structures}A), which may reflect the limitations of MLIPs in describing soft modes or numerical noise in the DFT forces used to construct the force constants.
The models therefore capture the vibrational response near equilibrium for the main spectral features.

These mobile-ion properties remain robust despite large atomic force errors, partly because the errors on framework atoms cancel within each polyhedron.
The properties examined above are largely governed by the motion of the mobile Na and Li ions, so we examine the force errors of the constituent \ce{VO6} octahedra and \ce{PO4} tetrahedra (Fig.~\ref{fig:cathode-property-comparison}E).
Across the evaluated NVP dataset, the overall atomic component RMSE is 0.173~eV/\AA.
For the \ce{VO6} framework units, the mean polyhedral net force RMSE is 0.029~eV/\AA, while the mean atomic force RMSE is 0.130~eV/\AA.
For the \ce{PO4} polyanions, the corresponding mean polyhedral net force and atomic force RMSEs are 0.041 and 0.146~eV/\AA, respectively.
Because vector summation can reduce the net residual even for randomly signed errors, we construct a random-sign baseline for reference, with means of 0.0455 and 0.0590~eV/\AA\ for \ce{VO6} and \ce{PO4}, respectively.
The observed mean net residuals are 37.4\% and 30.5\% below this baseline, indicating additional cancellation beyond the intrinsic reduction from vector summation.
A similar pattern is observed for the \ce{CoO6} units in \ce{Li_xCoO2}, where the mean net residual of 0.033~eV/\AA\ lies 25\% below the random-sign baseline of 0.044~eV/\AA\ (fig.~S13).
The force errors on individual framework atoms are therefore largely compensating rather than accumulating, as illustrated schematically in Fig.~\ref{fig:cathode-property-comparison}F.
A possible source of these atomic force errors is orbital information unresolved by the model inputs, since self-consistent DFT solutions with similar local magnetic moments can differ in orbital occupations \cite{ponet_energy_2024}.
When these electronic states are indistinguishable from the model inputs, the MLIP must represent their force labels through a single response, which approximates a sampling-weighted average when several states are represented in the training data, or the surface of the single state when only one is sampled.
Related work on thermal transport also shows that agreement in macroscopic observables can arise from compensation of errors in microscopic phonon contributions \cite{pota_thermal_2026}.

Structural relaxation from off-equilibrium configurations is where the geometry-only model departs most clearly from DFT (Fig.~\ref{fig:relaxed-structures}).
Detailed settings, including cell optimization and magnetic-moment initialization and optimization, are described in Methods.
For NVP, we select 15 structures containing V sites within $\mathrm{PC1}\in[-2,2]$ in Fig.~\ref{fig:fig1-summary}C.
The analysis comprises 120 V atoms spanning seven compositions between $x=1.25$ and 3, with each relaxed local environment characterized by the mean length of its six V--O bonds.
Fig.~\ref{fig:relaxed-structures}B compares these DFT-relaxed bond lengths with the MACE and mMACE predictions.
After DFT relaxation, the local magnetic moments are closely associated with the mean V--O bond lengths, consistent with the branches in Fig.~\ref{fig:fig1-summary}C.
Sites with moments of approximately 1--1.2~$\mu_\mathrm{B}$ have shorter bonds around 1.95~\AA, whereas those with moments of approximately 1.8--2~$\mu_\mathrm{B}$ have longer bonds around 2.03~\AA, consistent with V(IV) and V(III), respectively.
MACE broadly retains the two bond-length populations, but some low-moment sites relax to bond lengths characteristic of the longer-bond population, while some high-moment sites show the opposite deviation.
These cross-population discrepancies are consistent with the difficulty of distinguishing the overlapping magnetic branches identified above.
With explicit magnetic input, mMACE brings the mean V--O bond lengths of high-moment sites closer to the DFT values and better preserves the distinction between the magnetic-state-dependent structural branches.
The overall mean-bond-length MAE decreases from 8.7 to 5.3~m\AA, although some local environments remain imperfectly reproduced.
MACE, in contrast, reproduces the \ce{VO6} distortion better than mMACE, particularly for V(IV) octahedra (fig.~S14).
Remaining discrepancies may also reflect differences in relaxation algorithms and limited training-data coverage of the relevant structural and magnetic configurations.

For \ce{Li_xCoO2}, we relax ten structures at each of $x=0.5$ and 1 (Fig.~\ref{fig:relaxed-structures}C).
The DFT-relaxed structures again show a correspondence between local magnetic moments and mean Co--O bond lengths, whereas geometry-only MACE does not fully retain this magnetic-state-dependent structural variation.
For \ce{LiCoO2}, MACE predicts nearly uniform mean Co--O bond lengths of approximately 1.922~\AA.
This behavior can be understood from the training magnetic-moment distribution, which is dominated by low-moment Co sites, and is consistent with a learned effective surface biased toward the more frequently sampled low-spin-like environments.
With mMACE, the overall mean-bond-length MAE decreases from 10.8 to 8.0~m\AA.
Although this reduction is modest, mMACE recovers longer bonds for a subset of high-moment sites and brings them closer to the DFT high-moment branch. It reduces the absolute error in the mean Co--O bond length at 68.8\% of Co sites, including 82.9\% in \ce{LiCoO2} and 47.5\% in \ce{Li_{0.5}CoO2}.

The mMACE relaxations above start from the magnetic moments optimized by corresponding DFT relaxations.
In \ce{LiCoO2}, initializing every Co moment at a uniform $0.6~\mu_\mathrm{B}$ in both the DFT and mMACE relaxations instead leads to low-moment structures. Against these DFT-relaxed structures, geometry-only MACE has a mean-bond-length MAE of 3.5~m\AA, compared with 8.0~m\AA\ for the same \ce{LiCoO2} subset under the original magnetic initialization (fig.~S15), consistent with its learned effective surface favoring low-moment environments.
Geometry-only MACE cannot respond to a change in magnetic initialization at a fixed starting geometry, since its inputs contain only atomic species and positions.
When the intended calculation requires distinguishing and following magnetic-state-dependent relaxation branches, the model must therefore include information that distinguishes those states.

These comparisons separate the accuracy of selected physical properties from the preservation of magnetic-state-dependent local structures.
A geometry-only effective surface can retain useful mobile-ion migration landscapes while failing to reproduce the structural distinctions between magnetic branches, and magnetic models partly recover these distinctions in the present tests.

\section{Discussion}
\begin{figure}[tbp]
\centering
\includegraphics[width=\textwidth]{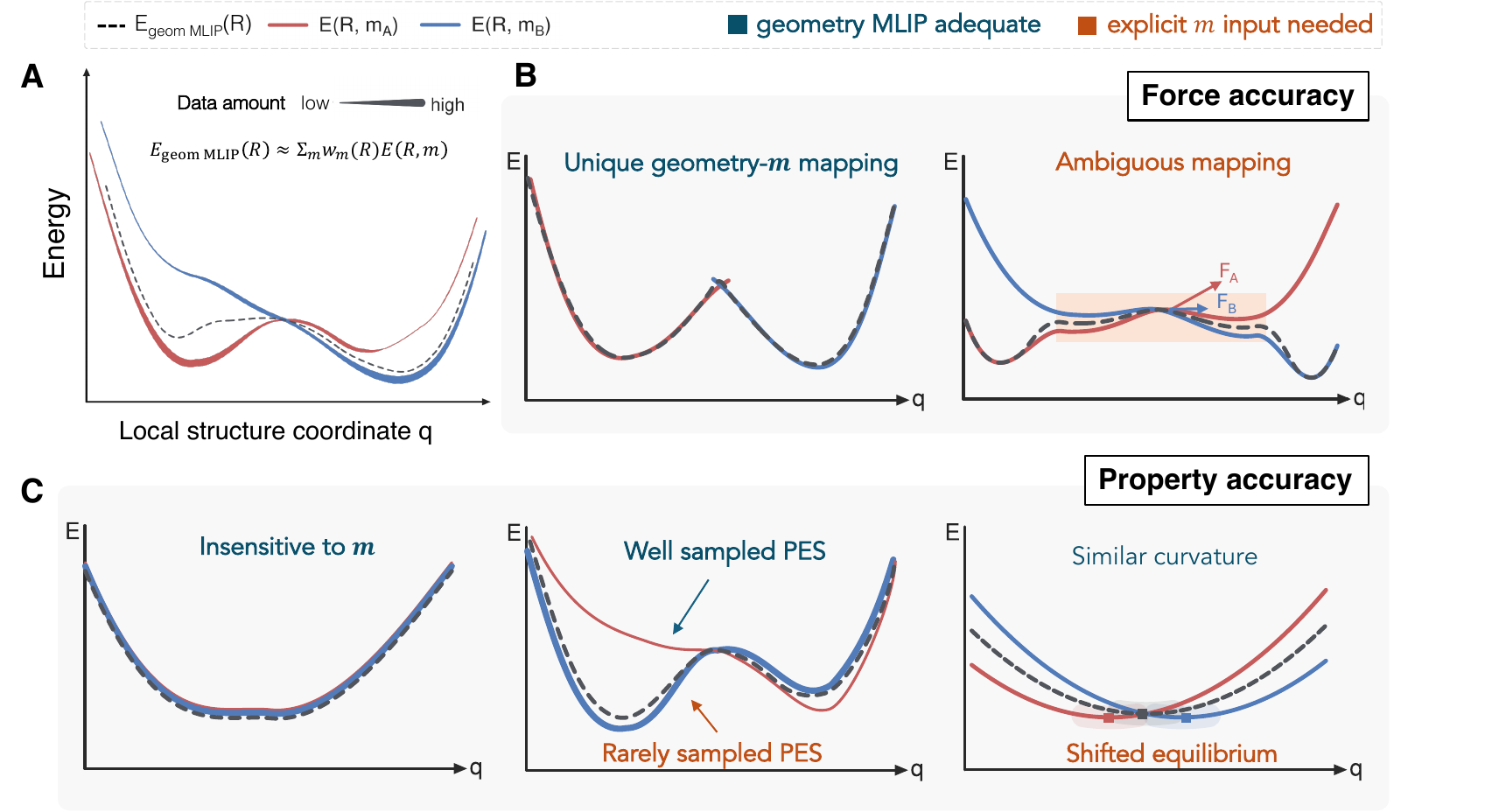}
\caption{\textbf{How magnetic input affects force and property predictions of MLIPs.}
(\textbf{A}) Influence of training-data sampling on the effective potential-energy surface (PES) learned by a geometry-only MLIP. Red and blue curves represent the magnetic branches $E(\mathbf{R},m_A)$ and $E(\mathbf{R},m_B)$, and the gray dashed curve represents the learned PES. Line thickness indicates the amount of training data along each branch, with $w_m(\mathbf{R})$ its schematic sampling weight, and $q$ is a structural coordinate.
(\textbf{B}) Need for explicit magnetic input for force accuracy, when geometry distinguishes the sampled magnetic states (left) and when similar geometries support distinct magnetic states (right). Shading marks the region of magnetic ambiguity. Arrows $F_A$ and $F_B$ illustrate different branch gradients at the same coordinate, even where the branch energies are similar.
(\textbf{C}) Need for explicit magnetic input for property accuracy. In the left panel, the relevant PES features depend weakly on the magnetic state. In the middle panel, the effective surface follows a well-sampled branch but deviates from a rarely sampled one. In the right panel, similar local curvatures coexist with shifted equilibrium positions. Blue and orange annotations indicate targets for which geometry-only input is adequate and those that require explicit magnetic information, respectively.
}
\label{fig:magnetic-model-assessment}
\end{figure}

Machine-learning interatomic potentials are increasingly trained on large DFT datasets in which many compounds are magnetic. This is especially true for foundational MLIPs, yet most of these models see only atomic positions.
We have shown that this becomes a problem where the geometry does not determine the magnetic state, and that the problem can be detected in the training data before any model is trained.
Force errors are large exactly where similar geometries support distinct magnetic states, a dataset-level descriptor of this magnetic ambiguity predicts the benefit of explicit magnetic input, and the consequences for physical properties depend on which features of the energy landscape a property relies on.
Fig.~\ref{fig:magnetic-model-assessment} brings these observations into a common picture.
The scenarios discussed here all assume adequate training data and converged model performance.

A geometry-only MLIP assigns a single energy and force to each atomic configuration.
In magnetic materials, a fixed geometry can admit several self-consistent DFT solutions, that is, distinct electronic states to which the calculation can converge \cite{burton_energy_2022}.
These solutions define distinct potential-energy surfaces (PESs) $E(\mathbf{R},m)$ that coexist over a range of geometries.
As plotted in Fig.~\ref{fig:magnetic-model-assessment}A, the PES learned by a geometry-only model can be represented schematically as their sampling-weighted combination,
\begin{equation}
    E_{\mathrm{geom\,MLIP}}(\mathbf{R})
    \approx\sum_m w_m(\mathbf{R})E(\mathbf{R},m),
    \label{eq:effective-geometry-pes}
\end{equation}
where $\mathbf{R}$ denotes the atomic coordinates at fixed composition, $m$ denotes the magnetic configuration, and $w_m(\mathbf{R})$ represents the relative sampling weight in the training data.
This picture is consistent with the \ce{LiCoO2} results, where low-moment Co sites dominate the training data and MACE relaxes even high-moment sites toward bond lengths characteristic of the low-moment population (Fig.~\ref{fig:relaxed-structures}C).
In practice, the learned PES also depends on the relative weights assigned to energy and force errors during training.

Which magnetic states enter the training data is decided by the DFT calculations, in which the magnetic state is relaxed self-consistently at each fixed geometry.
When DFT consistently converges to one magnetic branch, that branch dominates the sampled reference data, and a geometry-only MLIP can reproduce it closely.
When competing magnetic solutions are close in energy, the converged state can depend on calculation settings such as the initial magnetic moments, the starting wavefunction (table~S5) and the electronic minimization algorithm (fig.~S16).
These choices strongly affect the reference labels, as listed in Table~\ref{tab:dft-settings}.
Recomputing all 2,600 NVP training configurations at identical geometries with blocked-Davidson instead of conjugate-gradient minimization changes the force labels by 141~meV/\AA\ (RMSE), and additionally switching from Gaussian to tetrahedron smearing changes them by 153~meV/\AA, while the energy RMSE remains 2.2--2.5~meV/atom (Table~\ref{tab:dft-settings}, fig.~S16 and Sections~S2.2 and S5).
Both relabelled datasets are also easier to fit, with the MACE validation force RMSE decreasing from 204.7 to 175.0 and 187.0~meV/\AA\ as $T_{8\%}$ decreases from 0.86 to 0.67 and 0.79~$\mu_\mathrm{B}$.
However, as we discussed in the Results section, 
the force accuracy discussed here reflects how learnable the supplied labels are for a geometry-only MLIP, not their physical correctness, and DFT settings should be chosen on physical grounds, such as reaching the lowest-energy self-consistent solution, rather than by the resulting MLIP error.

\begin{table}[htbp]\centering\small
\caption{\textbf{Effect of DFT settings on the NVP dataset.} All settings use identical geometries. $\Delta E_{\mathrm{RMSE}}$ and $\Delta F_{\mathrm{RMSE}}$ are the per-atom energy RMSE and the force RMSE between the labels of each setting and those of the production setting. RMSE$_E$ and RMSE$_F$ are the validation errors after fine-tuning from MACE-MP-0. The preconditioned conjugate-gradient (CG) and blocked-Davidson algorithms were selected using \(\texttt{ALGO=All}\) and \(\texttt{ALGO=Normal}\) in VASP, respectively.}
\label{tab:dft-settings}
\begin{tabular}{@{}llccccc@{}}\toprule
Smearing & Algorithm & $\Delta E_{\mathrm{RMSE}}$ & $\Delta F_{\mathrm{RMSE}}$ & $T_{8\%}$ & RMSE$_E$ & RMSE$_F$ \\
 & & (meV/atom) & (meV/\AA) & ($\mu_\mathrm{B}$) & (meV/atom) & (meV/\AA) \\
 \midrule
Gaussian             & Preconditioned CG & --   & --    & 0.86 & 3.0 & 204.7 \\
Gaussian             & Blocked-Davidson        & 2.51 & 140.6 & 0.67 & 2.9 & 175.0 \\
Tetrahedron (Bl{\"o}chl) & Blocked-Davidson     & 2.18 & 153.2 & 0.79 & 3.2 & 187.0 \\
\bottomrule
\end{tabular}
\end{table}

The consequences for force prediction depend on how strongly the magnetic branches overlap in structure (Fig.~\ref{fig:magnetic-model-assessment}B).
In the left schematic, geometry-only models remain effective because the magnetic states are distinguishable through structural features, such as the Jahn--Teller distortions in \ce{Li_xMn2O4}, or because a preferred spin state dominates, as in \ce{CrO2}.
In the right schematic, similar geometries support distinct magnetic states.
Their PES branches can be nearly degenerate while retaining different gradients, and hence different forces $F_A$ and $F_B$, with $F_q=-\partial E/\partial q$.
A geometry-only model assigns a single force response to a given geometry and cannot resolve these competing force labels, whereas explicit magnetic information allows the branches to be distinguished.
The descriptor $T_p$ quantifies this magnetic ambiguity for a given dataset and provides an \textit{a priori} estimate of the potential benefit of magnetic input.
In this work, large $T_p$ values arise in three practical situations.

The first is electron localization in the DFT calculations.
The Hubbard $U$ correction, which favors localized electronic states by penalizing fractional orbital occupations \cite{dudarev_electron-energy-loss_1998}, can increase the number of metastable self-consistent solutions and change their relative energies \cite{ponet_energy_2024}.
This is consistent with the NVP results in Fig.~\ref{fig:fig1-summary}C, where similar local environments exhibit distinct moment amplitudes in the region where the two branches overlap.
The choice of functional has a similar effect.
In MatPES, the higher ambiguity of r$^2$SCAN relative to PBE involves both a larger magnetic-moment scale and a greater sign-related increment (fig.~S11), possibly because SCAN-family functionals predict larger moments \cite{fu_applicability_2018,ekholm_assessing_2018,kothakonda_testing_2023}, which amplify both amplitude differences and opposite-sign contrasts between structurally similar sites.

The second is the range of local magnetic moments permitted by the electronic configuration.
In MatPES, $T_{8\%}$ peaks at group 7 and is generally lower for $4d$ than for $3d$ elements.
Near-half-filled $d$ shells permit a wider range of signed magnetic moments and hence more distinct magnetic states in geometrically similar environments.

The third is structural sampling, which affects both the accessibility of magnetic states and their distinguishability from local geometry.
Through the coupling between magnetic states and lattice distortions \cite{moruzzi_ferromagnetic_1986,kormann_temperature_2014}, higher sampling temperatures broaden the distribution of local distortions and can change which self-consistent magnetic solutions are recorded in the subsequent DFT calculations.
The contrasting temperature dependences of \ce{N2VP} and \ce{Co3O4} show that this effect is material dependent, consistent with differences in their spin--lattice coupling.
Magnetic ambiguity is therefore also a useful diagnostic for assessing dataset construction before model training.

For physical properties, the question is whether the effective surface retains the features that govern the target observable (Fig.~\ref{fig:magnetic-model-assessment}C).
The left schematic illustrates a target coordinate along which the PES is insensitive to the magnetic state, such as the Li- and Na-ion migration coordinates examined in this work.
A geometry-only model can then retain the target property, partly through the cancellation of force errors within the coordinating environment.
This robustness is not universal.
Ahn et al. found that local hole-polaron formation in delithiated \ce{LiCoO2} changes the Li-migration barrier from approximately 0.21 to 0.34~eV \cite{ahn_first-principles_2023}, so agreement with one reference pathway does not establish that a model can describe coupled ion--polaron migration.
In the middle schematic, the effective PES follows a well-sampled magnetic branch but deviates from a less frequently sampled one, and property accuracy then depends on the magnetic state of interest.
In this work, for example, the geometry-only model captures the relaxed structures associated with low-moment Co sites more accurately than those associated with high-moment Co sites.
In the right schematic, the effective PES retains a local curvature similar to those of the magnetic branches, but its minimum lies at a different structural coordinate.
Comparable harmonic responses, such as similar phonon spectra, can therefore coexist with an inaccurate equilibrium structure for a specified magnetic state.

Omitting magnetic information therefore affects different properties differently.
The effective PES learned by a geometry-only MLIP may preserve relative energies or local curvature while shifting minima or losing the distinctions between magnetic branches.
Beyond the cases examined here, explicit magnetic information is also necessary to distinguish magnetic excited states at a fixed geometry, as in a paramagnetic-to-ferromagnetic transition.
It may also be important for defects that support competing charge-localization or spin states when their associated lattice distortions do not sufficiently distinguish these states.
More generally, the same analysis could be applied to other electronic quantities absent from the model input, such as atomic charges, orbital occupations, or the local spin and charge populations that distinguish excited states, by replacing the local moment in $a_i$ with the corresponding per-atom quantity.

The need for explicit magnetic information therefore depends on both the magnetic ambiguity of the training data and the sensitivity of the target properties to the resulting PES differences.
The descriptor $T_p$ connects dataset magnetic ambiguity to the force-fitting benefit of explicit magnetic information across material-specific datasets and MatPES, and it can be evaluated before model training from quantities that spin-polarized DFT calculations already provide.
This assessment concerns the learnability of the reference data, while the suitability of a geometry-only model for an application depends on whether its effective surface retains the relevant PES features.
Explicit magnetic information becomes particularly important when the objective is to resolve competing force responses or to select and follow distinct magnetic-state-dependent branches.
Combining dataset-level assessment with validation of the target properties provides a practical basis for choosing an MLIP description.
For the large datasets now used to train foundation models across battery, catalysis and magnetic-materials research, this means that the decision to store and learn magnetic degrees of freedom can be made from the data, element by element, and that DFT settings should be judged by the physical states they produce rather than by how easily the resulting labels are fitted.

\section{Methods}

\subsection{DFT calculations}

DFT calculations were performed using the Vienna ab initio Simulation Package (VASP), version 6.3.2 \cite{kresse_efficiency_1996,kresse_efficient_1996}.
The exchange-correlation energy was approximated by the Perdew--Burke--Ernzerhof (PBE) generalized gradient approximation \cite{perdew_generalized_1996}.
Wave functions were expanded in a plane-wave basis set truncated at a kinetic energy cutoff of 520~eV, and projector augmented-wave (PAW) potentials were employed to describe core--valence electron interactions \cite{kresse_ultrasoft_1999}.
The PAW (PBE\_54) potentials used were \texttt{Na\_pv} (19 Sep 2006), \texttt{Li\_sv} (10 Sep 2004), \texttt{V} (07 Sep 2000), \texttt{Cr} (06 Sep 2000), \texttt{Co} (02 Aug 2007), \texttt{P} (06 Sep 2000) and \texttt{O} (08 Apr 2002).
The first Brillouin zone was integrated using $\Gamma$-centered Monkhorst--Pack sampling with a spacing of approximately 0.2~\AA$^{-1}$ in each lattice direction.
Total energies were converged within $10^{-5}$~eV/cell during the electronic minimization, without preserving any symmetry.
The Hubbard $U$ correction \cite{dudarev_electron-energy-loss_1998} was applied following the MITRelaxSet settings in \texttt{pymatgen}, with $U$ values of 3.1, 3.5 and 3.4~eV for V, Cr and Co, respectively, and an additional exchange parameter of $J=0.9$~eV for Cr.
Datasets without the Hubbard correction were computed with $U=0$ and otherwise identical settings.

Spin-polarized calculations were initialized with V moments set to approximately $1~\mu_\mathrm{B}$, and Co moments to $0.6~\mu_\mathrm{B}$, and used conjugate-gradient electronic minimization (\texttt{ALGO = All}) with Gaussian smearing (\texttt{ISMEAR = 0}, \texttt{SIGMA =} 0.05) and \texttt{LMAXMIX = 4}.
Local magnetic moments were obtained using the PAW-projector-based site decomposition implemented in VASP.
Non-spin-polarized reference data were computed at the same geometries with \texttt{ISPIN = 1} and otherwise identical settings.
The alternative electronic-minimization settings used to recompute the NVP training configurations are listed in Table~\ref{tab:dft-settings}.

For the fixed-geometry calculations in fig.~S4, one V atom in the ground-state \ce{N3VP} structure was displaced by 0.25~\AA.
Each structure contains eight V atoms. Ferromagnetic initializations assigned $+2~\mu_\mathrm{B}$ to all eight V sites, while antiferromagnetic initializations assigned $-2~\mu_\mathrm{B}$ to four sites and $+2~\mu_\mathrm{B}$ to the other four.

\subsection{Dataset construction}

Reference configurations for the material-specific datasets were generated following the workflow of Deng et al. \cite{deng_accelerating_2025}.
MD simulations were performed using MACE-MP-0a at 800~K for 15~ps with a time step of 1~fs. After discarding the first 5~ps for equilibration, configurations were randomly selected from the remaining trajectories and subjected to random strains within $\pm3\%$. Single-point DFT calculations were then performed using the settings described above. The number of configurations in each dataset is listed in table~S1.
For the temperature-dependent ambiguity analysis in fig.~S8, the same sampling procedure was repeated at 50, 100, 200, 300, 400, 500, 600, 700, 800 and 900~K.

\subsection{Model training}

All material-specific MACE models were fine-tuned from the medium-size MACE-MP-0 foundation model pre-trained on the MPtrj dataset \cite{batatia_foundation_2025}, using multi-head fine-tuning.
Each dataset was randomly split into training and validation sets with a validation fraction of 0.2.
Models were trained with a learning rate of 0.01 for at most 500 epochs, with early stopping if the validation loss did not improve for 100 consecutive epochs, and a batch size of 32.
The energy, force and stress loss weights were 1, 10 and 1, respectively. 
The radial cutoff was 6~\AA, inherited from the foundation model.
The mMACE models were fine-tuned from an mMACE model pre-trained on the MatPES-PBE dataset, with all other settings identical to those above.

For MatPES-PBE and MatPES-r$^2$SCAN \cite{kaplan_foundational_2025}, we trained the MACE and mMACE models from scratch using the training settings of Ho et al. \cite{ho_equivariant_2026}.
The models used the same training and validation datasets, with two interaction layers, hidden representations of $128\times0e+128\times1o$, a correlation order of 3, a radial cutoff of 6~\AA, eight radial basis functions, and a maximum spherical-harmonic degree of 3. The mMACE model additionally incorporated local DFT magnetic moments, with eight magnetic radial basis functions and a maximum magnetic angular degree of 1.
Training used the Adam optimizer with AMSGrad, an initial learning rate of 0.005, exponential moving averaging, and energy, force, and stress loss weights of 1, 10, and 5, respectively. Models were trained in double precision for up to 300 epochs, with an early-stopping patience of 40 epochs. MACE and mMACE were trained on two and four NVIDIA B200 GPUs, respectively, with a batch size of 300 configurations per GPU, giving total batch sizes of 600 and 1200.
Configurations with reference forces above 20~eV/\AA\ were excluded from both training and the element-resolved analysis.

The CHGNet model used for comparison on the NVP dataset at $U=3.1$~eV was fine-tuned from the CHGNet 0.3.0 universal model, pretrained on the MPtrj dataset \cite{deng_chgnet_2023}.

\subsection{Dataset-level magnetic ambiguity}
\label{sec:ambiguity-methods}

For an atomic site $i$ with DFT reference moment $m_i$, its local environment is represented using periodic SOAP descriptors implemented in DScribe \cite{himanen_dscribe_2020}, with $r_{\mathrm{cut}}=6$~\AA, $n_{\max}=8$, $l_{\max}=3$ and a Gaussian width of 1~\AA.
The descriptors retain species-pair channels and include all atomic species in the surrounding environment.
Each raw SOAP vector $\mathbf{s}_i$ is L2-normalized,
\begin{equation}
    \hat{\mathbf{x}}_i=\frac{\mathbf{s}_i}{\|\mathbf{s}_i\|_2}.
    \label{eq:soap-normalisation}
\end{equation}
The distance between sites is the Euclidean distance between their normalized SOAP vectors,
\begin{equation}
    d_{ij}=\sqrt{\max\left(2-2\hat{\mathbf{x}}_i\cdot\hat{\mathbf{x}}_j,0\right)}.
    \label{eq:soap-distance}
\end{equation}

Exact nearest-neighbor searches are performed over the full same-element population, excluding only the query site itself.
The 100 nearest candidates are sorted by distance to form a set $K_i$, and the fifth-nearest distance, $d_{i,5}$, is used to determine the adaptive neighborhood as
\begin{equation}
    N_i=\left\{j\in K_i:d_{ij}\leq \delta_i \right\},\qquad
    \delta_i = 1.25\,d_{i,5},\qquad 5\leq|N_i|\leq100.
    \label{eq:adaptive-neighborhood}
\end{equation}

For each element, we reverse the reference moments $m_i$ of all atoms within a DFT configuration whenever the net moment of that element is negative, and denote the resulting sign-fixed moments by $\tilde{m}_i$.
If the net moment vanishes, the sign is chosen such that the largest-magnitude moment of that element is positive.
This convention is used only to compute the descriptor.
With these neighbor sites, the site ambiguity $a_i$ of atom $i$ is defined as
\begin{equation}
    a_i=\operatorname{median}_{j\in N_i}|\tilde{m}_i-\tilde{m}_j|.
    \label{eq:site-ambiguity}
\end{equation}
The median limits the influence of unusually large outliers.

For a dataset with $N$ target sites, the ambiguities are ranked as $a_{(1)}\geq a_{(2)}\geq\cdots\geq a_{(N)}$.
The dataset descriptor $T_p$, in units of $\mu_\mathrm{B}$, is the mean over the fraction $p$ of sites with the largest site ambiguities,
\begin{equation}
    T_p=\frac{\displaystyle\sum_{r=1}^{\lfloor Np \rfloor}a_{(r)}}{\lfloor Np \rfloor}.
    \label{eq:tail-ambiguity}
\end{equation}

For a pair of sign-aligned moments, the absolute difference can be written as
\begin{equation}
|\tilde m_i-\tilde m_j|
=\big||\tilde m_i|-|\tilde m_j|\big|
+2\min(|\tilde m_i|,|\tilde m_j|)\,\mathbf{1}_{\tilde m_i\tilde m_j<0}.
\label{eq:moment-difference-decomposition}
\end{equation}
We compute the magnitude-only site ambiguity as the median of the first term over the same neighbors $N_i$.
To decompose $T_p$, we average these magnitude-only scores over the sites selected by the signed scores $a_i$, giving $T_p^{\mathrm{mag}}$.
The difference $\Delta T_p^{\mathrm{sign}}=T_p-T_p^{\mathrm{mag}}$ gives the additional ambiguity associated with retaining the signs on that same set of sites,
\begin{equation}
T_p=T_p^{\mathrm{mag}}+\Delta T_p^{\mathrm{sign}}.
\label{eq:tail-sign-increment}
\end{equation}

For each site of the nonmetals N, C, P, O, S, Se, F, Cl, Br and I in MatPES, we assign a site ambiguity equal to the mean site ambiguity of its directly bonded $d$- and $f$-block neighbors. Bonds are identified using a distance cutoff of 1.2 times the sum of the two atoms' covalent radii tabulated in ASE. We then calculate $T_{8\%}$ from these nonmetal-site scores using the same upper-tail average, while $I_F$ is evaluated from forces on the nonmetal atoms themselves.

Sensitivity tests for the adaptive-distance multiplier, fixed neighbor count and upper-tail fraction are provided in fig.~S5.
The descriptor also requires a sufficiently large site population.
Subsampling the five material-specific datasets shows that, from about 1,000 central-element sites onward, the median deviation of $T_{8\%}$ from its full-dataset value is below 15\% (fig.~S6).
We therefore recommend evaluating $T_p$ on populations of at least 1,000 sites.

For each population, force MAEs are calculated over identical central-element sites for the matched MACE and mMACE models,
\begin{equation}
    \mathrm{MAE}_{M}=\frac{1}{3N}\sum_{i=1}^{N}\sum_{\alpha=x,y,z}
    \left|F^{M}_{i\alpha}-F^{\mathrm{DFT}}_{i\alpha}\right|,
    \label{eq:central-force-mae}
\end{equation}
where $M$ denotes the model and $F_{i\alpha}$ is a Cartesian force component.
The relative improvement in force accuracy, $I_F$, is calculated from the force MAEs as
\begin{equation}
    I_F=100\left(1-\frac{\mathrm{MAE}_{\mathrm{mMACE}}}{\mathrm{MAE}_{\mathrm{MACE}}}\right)\%,
    \label{eq:force-mae-reduction}
\end{equation}
so positive values indicate smaller force errors with mMACE.

The SOAP power spectrum is not a complete description of the local environment, and environments that MACE separates through message passing can lie close together in SOAP space.
Such environments are counted as neighbors, and their moment differences raise $a_i$, so $T_p$ tends to overestimate the ambiguity seen by a more complete representation.

\subsection{Ambiguity threshold in MatPES}
\label{sec:threshold-methods}

For the element populations of each MatPES functional, the dependence of $I_F$ on $T_{8\%}$ is described by a constant baseline followed by a linear rise,
\begin{equation}
    I_F=b_0+s\,\max\left(0,\;T_{8\%}-T^{*}\right),
    \label{eq:threshold-fit}
\end{equation}
where $T^{*}$ is the threshold, $b_0$ the baseline and $s$ the slope above the threshold.
The parameters are obtained by minimizing the sum of absolute deviations over all non-lanthanide elements, which limits the influence of individual outlying elements.
$T^{*}$ is scanned from 0.3 to 4.0~$\mu_\mathrm{B}$ in steps of 0.01~$\mu_\mathrm{B}$, with $b_0$ and $s$ optimized at each value.
For MatPES-PBE, we retain $T^{*}=1.25~\mu_\mathrm{B}$ for consistency with the preceding analysis, whereas for MatPES-r$^2$SCAN, the fitted minimum lies at $T^{*}=0.93~\mu_\mathrm{B}$.
With $T^{*}$ fixed, the line above the threshold is a separate least-absolute-deviation linear fit to the non-lanthanide elements with $T_{8\%}\geq T^{*}$.

\subsection{Structural relaxation}
\label{sec:mmace-spin-relaxation}

DFT structural relaxations optimized both atomic positions and cell parameters using the conjugate-gradient algorithm, with a force convergence threshold of $0.01$~eV/\AA\ and a maximum of 300 ionic steps. 
MACE and mMACE relaxations used the LBFGS optimizer in ASE with a \texttt{FrechetCellFilter} at zero external pressure and the same convergence threshold.
During mMACE relaxation, the local magnetic moments were optimized using the calculator's self-consistent spin optimization procedure.
For each structure, the initial local magnetic moments were taken from the corresponding DFT-relaxed structures.
The spin optimization allowed up to 1000 iterations, with a step size of $10^{-3}$  and a convergence tolerance of $10^{-4}$.
In the initialization test of fig.~S15, the DFT and mMACE relaxations of \ce{LiCoO2} were instead initialized to the default \texttt{pymatgen} Co moment of \(0.6~\mu_\mathrm{B}\), with otherwise identical settings.

\subsection{Polyhedral force errors}

For each \ce{VO6} octahedron or \ce{PO4} tetrahedron $P$ containing $N_P$ atoms, the atomic force RMSE is
$\mathrm{RMSE}^{\mathrm{atom}}_P=\sqrt{\frac{1}{3N_P}\sum_{i\in P}|\Delta\mathbf{F}_i|^2}$ and the polyhedral net force residual is $\Delta\mathbf{F}_P=\frac{1}{N_P}\sum_{i\in P}\Delta\mathbf{F}_i$, with $\mathrm{RMSE}^{\mathrm{net}}_P=|\Delta\mathbf{F}_P|/\sqrt{3}$.
The reported values are means over all polyhedra in the evaluated NVP dataset.
For the random-sign baseline, the sign of each atomic force-error vector $\Delta\mathbf{F}_i$ is reversed independently, and the net residual is averaged over all $2^{N_P}$ sign combinations and all polyhedra.
\subsection{Voltage profile}

The voltage profile of \ce{Na_xV2(PO4)3} was obtained from the calculated energies of the stored configurations at each composition $x$.
For each model, the lowest energy per formula unit at each $x$, $E(x)$, was referenced to the line joining the \ce{N1VP} and \ce{N4VP} endpoints to construct the convex hull of formation energies, and the compositions on the lower hull were taken as the stable phases.
The average voltage between two adjacent stable compositions $x_1<x_2$ is \cite{aydinol_ab_1997}
\begin{equation}
V(x_1\rightarrow x_2) = -\frac{E(x_2)-E(x_1)-(x_2-x_1)\,\mu_{\mathrm{Na}}}{(x_2-x_1)\,e},
\label{eq:voltage}
\end{equation}
where $\mu_{\mathrm{Na}}$ is the energy per Na atom in the reference state, and $e$ is the elementary charge.
Taking fully sodiated \ce{N4VP} as the starting point, the voltage for extracting $(4-x)$ Na per formula unit down to \ce{N_xVP} is $V=-[E(4)-E(x)-(4-x)\mu_{\mathrm{Na}}]/[(4-x)e]$, and the step-wise profile in Fig.~\ref{fig:cathode-property-comparison}A is obtained by applying Eq.~\ref{eq:voltage} to consecutive lower-hull vertices.
The same $\mu_{\mathrm{Na}}$ is used for all segments of a given model, so it shifts the entire profile by a constant.
Because each calculated profile is aligned to the first experimental voltage point, the comparison between models and experiment depends only on the relative energies of the stable compositions.

\subsection{MD simulations}

MD simulations for ionic transport were performed with ASE using $2\times2\times1$ supercells of \ce{N3VP} at 500, 600, 700, 800 and 900~K.
The simulations were run in the canonical (NVT) ensemble with a Berendsen thermostat \cite{berendsen_molecular_1984}.
The integration time step and thermostat coupling time were both 1~fs. 
Each simulation was run for 1~ns, with the initial 100~ps discarded for equilibration.

\subsection{Ionic transport properties}

The temperature-dependent mean squared displacement (MSD) of the mobile ions was calculated from their trajectories using the \texttt{kinisi} package \cite{mccluskey_accurate_2025},
\begin{equation}
\mathrm{MSD}(T, t) = \frac{1}{N}  \sum_{i=1}^{N} \left| \mathbf{R}_i(t) - \mathbf{R}_i(0) \right|^2,
\end{equation}
where $N$ is the number of mobile ions in the simulation cell and $\mathbf{R}_i(t)$ is the position of the $i$-th ion at time $t$.
The tracer diffusion coefficient $D(T)$ is extracted from the linear slope of the MSD as a function of time according to the Einstein relation,
\begin{equation}
D(T) = \frac{1}{2n} \lim_{t \to \infty} \frac{d}{dt} \mathrm{MSD}(T, t),
\end{equation}
where $n=3$ is the dimensionality of the Na diffusion considered here.
The temperature dependence of the diffusivity was fitted to an Arrhenius relationship to determine the activation energy $E_a$,
\begin{equation}
D(T) = D_0 e^{-E_a/(k_\mathrm{B} T)},
\end{equation}
where $D_0$ is the pre-exponential factor and $k_\mathrm{B}$ is the Boltzmann constant.
The uncertainties of $E_a$ were estimated from the standard deviation of this fit.
The ionic conductivity $\sigma(T)$ is related to the diffusion coefficient by the Nernst--Einstein equation,
\begin{equation}
\sigma(T) = \frac{N q^2}{V k_\mathrm{B} T H_R} D(T),
\end{equation}
where $V$ is the volume of the simulation cell, $q$ is the charge of the mobile ion and $H_R$ is the Haven ratio, which accounts for correlations between ionic motions and is set to $H_R=1$ here.

\subsection{Phonon calculations}

Phonon calculations were performed using the finite-displacement method implemented in Phonopy \cite{togo_implementation_2023}, with a $2\times 2\times 1$ supercell for each structure.
Symmetry-inequivalent displaced supercells were generated using the default displacement amplitude of 0.01~\AA, and the forces on the displaced structures were calculated with DFT, MACE at $U=0$ and $U=3.1$~eV, or mMACE. For mMACE, the local magnetic moments were fixed at the values obtained from the corresponding relaxed structure.

The resulting harmonic force constants were used to calculate the total and element-projected phonon densities of states on an $8\times8\times8$ q-point mesh using the tetrahedron method. 
For each comparison, the same initial structure was relaxed separately with each method before the phonon calculation.

\section*{Acknowledgments}
\paragraph*{Funding:}
Z.D. acknowledges the support from A*STAR under its MTC-YIRG (H24-MRG0889), NRF-Singapore under its Competitive Research Programme (NRF-CRP30-2023-0001), and the Ministry of Education, Singapore, under the Academic Research Fund Tier 1 (FY2026). B.Z. acknowledges the support by the National Natural Science Foundation of China (Grant No.12404068). Computational work involved in this research work is partially supported by NUS IT’s Research Computing group using grant number NUSREC-HPC-00001.

\paragraph*{Author contributions:}
Conceptualization: Z.D. Methodology: Y.D., C.H.H., B.Z., Z.D. Investigation: Y.D., M.L., L.L., B.Z., Z.D. Formal analysis: Y.D., L.L. Supervision: Z.D. Writing, original draft: Y.D., Z.D. Writing, review and editing: all authors.

\paragraph*{Competing interests:}
C.H.H. is currently employed as a research intern at Mistral AI. This work was conducted independently of Mistral AI, which provided no funding or resources for this research.

\paragraph*{Data, code and materials availability:}
The data supporting the findings of this study, including atomic configurations, corresponding atomic force errors, and magnetic ambiguity results, together with the code used to calculate magnetic ambiguity, are publicly available at \url{https://github.com/deng-group/magnetic_ambiguity}.

\paragraph*{Supplementary materials:}
Supplementary Text, Figs. S1 to S16, Tables S1 to S5.

\bibliographystyle{bibliography}
\bibliography{references}

\end{document}